\documentclass[12pt,preprint]{aastex}

\usepackage{graphicx, color}
\usepackage{dcolumn}
\usepackage{bm}
\usepackage{booktabs}
\usepackage{natbib}
\usepackage{lscape}

\newcommand {\dif}[3][]{\frac{d^{#1}#2}{d#3^{#1}}}
\newcommand {\pdif}[3][]{\frac{\partial^{#1}#2}{\partial#3^{#1}}}
\newcommand {\fr}{\frac}
\newcommand {\lsim}{\hspace{0.3em}\raisebox{0.4ex}{$<$}\hspace{-0.75em}\raisebox{-.7ex}{$\sim$}\hspace{0.3em}}
\newcommand {\gsim}{\hspace{0.3em}\raisebox{0.4ex}{$>$}\hspace{-0.75em}\raisebox{-.7ex}{$\sim$}\hspace{0.3em}}

\makeatletter
\def\mart{\@ifnextchar[{\mart@@}{\mart@}}
\def\mart@@[#1]#2{\sqrt[#1]{\mathstrut{#2}}}
\def\mart@#1{\sqrt{\mathstrut{#1}}}
\makeatother

\newcommand{\myemail}{minoshim@stelab.nagoya-u.ac.jp}
\newcommand{\Yohkoh}{\it Yohkoh}

\newcommand{\RHESSI}{\it RHESSI}
\newcommand{\FP}{Fokker-Planck}

\newcommand{\gyros}{gyrosynchrotron}

\begin{document}

\shorttitle{Numerical Study of Microwave Propagation in Flare}
\shortauthors{Minoshima and Yokoyama}

\title{Numerical Study of a Propagating Non-Thermal Microwave Feature in a Solar Flare Loop}

\author{T. Minoshima\altaffilmark{1,2} and T. Yokoyama\altaffilmark{1}}
\altaffiltext{1}{
Department of Earth and Planetary Science, Graduate School of Science, University of Tokyo,
7-3-1, Hongo, Bunkyo-ku, Tokyo, 113-0033, Japan;
}
\altaffiltext{2}{
Solar-Terrestrial Environment Laboratory, Nagoya University,
Furo-cho, Chikusa-ku, Nagoya 464-8601, Japan;
}
\email{\myemail}


\begin{abstract}
We analytically and numerically study the motion of electrons along a magnetic loop, to compare with the observation of the propagating feature of the non-thermal microwave source in the 1999 August 28 solar flare reported by Yokoyama et al. (2002). We model the electron motion with the Fokker-Planck equation and calculate the spatial distribution of the gyrosynchrotron radiation. We find that the microwave propagating feature does not correspond to the motion of electrons with a specific initial pitch angle.
This apparent propagating feature is a consequence of the motion of an ensemble of electrons with different initial pitch angles, which have different time and position to produce strong radiation in the loop.
 We conclude that the non-thermal electrons in the 1999 August 28 flare were isotropically accelerated and then were injected into the loop.
\end{abstract}


\section{Introduction}\label{sec:introduction}
In solar flares, a large amount of released energy is expended for production of non-thermal particles. They propagate in the solar corona and produce non-thermal emissions such as hard X-rays (HXRs), $\gamma$-rays, and microwaves. The observed non-thermal emissions have key information on acceleration and transport mechanisms of particles.

Flare non-thermal emissions are signatures of propagation of accelerated particles. For HXRs, \cite{1994PhDT.......335S} analyzed flares with double-footpoint HXR sources observed with {\Yohkoh} \citep{1991SoPh..136....1O}, and showed that the time variation of the emissions from double sources coincides with each other within 0.1 s. This indicates that electrons, not ions, propagate along the magnetic loop, reach the footpoints, and then emit HXRs there. 
\cite{2003ApJ...595L..69L} observed intense HXR and $\gamma$-ray emissions in the 2003 July 23 flare, by using {\it Reuven Ramaty High Energy Solar Spectroscopic Imager} \citep[{\RHESSI};][]{2002SoPh..210....3L}. They found that the location of the $\gamma$-ray line emission source, which corresponds to a precipitation site of accelerated ions, does not coincide with the HXR sources. This implies that the acceleration site and/or propagation path of ions differs from that of electrons.

For meter-wavelength radio emissions, the so-called type III radio burst \citep[e.g.,][]{1998ARA&A..36..131B} is thought to be due to electromagnetic conversion of Langmuir waves, which are excited by electrons propagating from a flare site toward the interplanetary space along open magnetic field lines.
 For centimeter-wavelength radio emissions, the microwave source located at a remote distance from a flare main site \citep[microwave remote source;][]{1996SoPh..165..275H,1997SoPh..173..319H,1997ApJ...489..976N} is interpreted as the precipitation of electrons propagating from the main site where the energy release and particle acceleration occur. \cite{1999PASJ...51..483H} showed that the light curve at the microwave remote source has a similar profile to the HXR light curve at the main site, but is slightly delayed by roughly the loop transit time of electrons.
{On the other hand, \cite{2008ApJ...677.1367A} reported a flare in which the microwave light curve slightly leads the HXR. They argued that this time delay, opposite to that found by \citeauthor{1999PASJ...51..483H}, could be realized if electrons are injected into an asymmetric magnetic loop toward a footpoint with a stronger magnetic field.}
 \cite{2008ApJ...673..598M} analyzed the spatial and spectral variations of HXR and microwave emissions in the 2003 May 29 flare, and showed that the observed characteristics can be interpreted as a consequence of the electron transport called trap-plus-precipitation \citep{1976MNRAS.176...15M}. 
 
There are a few microwave observations of the direct detection of propagating electrons from a time series of images. \cite{1994kofu.symp..199B} reported the motion of the 8.4 GHz and 15 GHz microwave sources with a speed of 3,000 ${\rm km \; s^{-1}}$. \cite{2000ApJ...533L.167W} reported the rapid motion of the 0.33 GHz microwave source with a speed of $2.6 \times 10^{4} \; {\rm km \; s^{-1}}$. The most striking observation was reported by \cite{2002ApJ...576L..87Y} (hereafter Y2002).
 They observed a flare that occurred on 1999 August 28, using Nobeyama Radioheliograph \cite[NoRH;][]{1994PROCIEEE...82..705} with 0.1 s cadence.
 From a time series of NoRH 17 GHz images, they found two different classes of propagating features of non-thermal microwave sources along the loop (see Fig. 3 in Y2002). One is a slower propagation at a speed of $6 \times 10^{3} \; {\rm km \; s^{-1}}$, and the other is a faster propagation at a speed of $9 \times 10^{4} \; {\rm km \; s^{-1}}$. 

 The origin of the slower propagating feature is discussed by \cite{2007A&A...465..613S}. They interpreted this feature as propagation of low-frequency whistler waves that can trap the parent electrons by wave-particle interactions at their turbulent front. The phase velocity of low-frequency whistler waves is roughly in agreement with the observation. The origin of the faster propagating feature is discussed by Y2002. They interpreted this propagating feature as a free stream of relativistic electrons along the loop. Assuming that the 17 GHz microwave-emitting electrons are almost relativistic ($v \sim c > 9 \times 10^{4} \; {\rm km \; s^{-1}}$), they concluded that the parent electrons are injected with large pitch angle, $\sim 70^{\circ}$.

The interpretation of Y2002 on the microwave propagating feature is subject to the following assumptions. First, they assumed that electrons freely stream along the loop. However, since the flare loop must be a converging magnetic loop, electrons suffer the magnetic mirroring force. 
Second, they assumed that the apparent motion of the microwave source, which is generated via {\gyros} radiation \citep{1969ApJ...158..753R,1972SoPh...26..151T,1981ApJ...251..727P,1985ARA&A..23..169D,1999spro.proc..211B}, corresponds to the trajectory of the parent electrons.
 However, since the {\gyros} radiation mechanism intricately depends on many physical parameters, it is not necessarily evident that the apparent motion of the radiation source is identical to the trajectory of the electrons with a specific pitch angle.

In this paper we reconsider the rapidly propagating feature of the microwave source in the 1999 August 28 flare and address the pitch-angle distribution of the electrons, by refining the treatments of the emission mechanism as well as the electron motion. In {\S}~\ref{sec:analytic-treatment} we present an analytic solution of the electron motion in a converging magnetic loop. {\S}~\ref{sec:numerical-model} shows our numerical model of the electron motion along the loop, in which we use the gyro-averaged Fokker-Planck equation to determine the electron phase space distribution. We calculate the {\gyros} intensity distribution along the loop from the calculated electron distribution for comparison with the observation. In {\S}~\ref{sec:result-discussion} we present our calculation results. The results do not support the interpretation of Y2002 but suggest that the electrons in the 1999 August 28 flare were isotropically accelerated and then were injected into the loop. In {\S}~\ref{sec:summary} we summarize this paper.

\section{Analytic Treatment}\label{sec:analytic-treatment}
 We show an analytic solution of the motion of a single electron in a converging magnetic loop that helps to understand calculation results in {\S} \ref{sec:result-discussion}. 
This is described by the following equations: the equations of motion along the magnetic field line and the conservation of the magnetic moment,
\begin{eqnarray}
\dif{r}{t} &=& \mu v, \label{eq:drdt}\\
\frac{1-\mu^{2}}{B(r)} &=& \frac{1-\mu_{0}^{2}}{B(0)}, \label{eq:mag_mom}
\end{eqnarray}
 where $r$ is the position measured from the loop top ($r = 0$ is the initial position) along the field line, $v$ is the velocity, $\mu$ is the pitch-angle cosine, $\mu_{0}$ is the initial pitch-angle cosine, and $B(r)$ is the magnetic field strength at the position, respectively. We do not consider the energy change of an electron, thus $v$ = constant.

For mathematical convenience, we give the magnetic field strength as a quadratic function with respect to $r$ of the form \cite[e.g.,][]{1996ApJ...464..985A,1999PASJ...51..483H}:
\begin{eqnarray}
B(r) &=& B_{0} + (B_{\rm f} - B_{0})\left(\frac{r}{r_{\rm f}}\right)^{2}
= B_{0} \left\{ 1 + (M-1)\left(\frac{r}{r_{\rm f}}\right)^{2} \right\},\label{eq:quadb}
\end{eqnarray}
where $B_{0}$ and $B_{\rm f}$ are the field strengths at the loop top and at the footpoint, $r_{\rm f}$ is the footpoint distance from the loop top along the field line (a half length of the loop), and $M = B_{\rm f}/B_{0}$ is the magnetic mirror ratio, respectively. When its initial pitch-angle cosine $\mu_0$ is smaller than the loss cone angle cosine $\mu_{\rm c} = \mart{1-M^{-1}}$, an electron is trapped and has a bounce motion in the loop due to the magnetic mirror. With a help of equation (\ref{eq:quadb}), equations (\ref{eq:drdt}) and (\ref{eq:mag_mom}) give a solution of the bounce motion as simple harmonic oscillation:
\begin{eqnarray}
 r &=& R \sin (\omega t), \label{eq:sol_r} \\
 \mu &=& \mu_{0} \cos (\omega t), \label{eq:sol_mu}
\end{eqnarray}
where
\begin{eqnarray}
R &=& \frac{\mu_{0}}{\mart{(1-\mu_{0}^{2})(M-1)}} r_{\rm f}, \label{eq:rr} \\
\omega &=& \mart{(1-\mu_{0}^{2})(M-1)}\frac{v}{r_{\rm f}}. \label{eq:omega}
\end{eqnarray}
Equation (\ref{eq:omega}) tells that an electron with a smaller $\mu_{0}$ (larger initial pitch angle) has a higher frequency of the bounce motion.

We apply the above solution to the propagating feature of the microwave source reported by Y2002.
 We assume, in the same manner as Y2002, that the apparent propagating motion of the radiation source corresponds to the trajectory of electrons with a specific initial pitch angle. The observation gives constraints on several variables appearing in equations (\ref{eq:rr}) and (\ref{eq:omega}). 
According to the analysis by Y2002 (see Fig. 3 in their paper), the electrons traveled $\sim 4.5 \times 10^{4} \; {\rm km}$ from the injection site (loop top) to the loop end (footpoint).
We adopt this value as the footpoint distance $r_{\rm f}$, and evaluate $R \sim r_{\rm f}$.
An observed travel time of 0.5 s is roughly equal to a quarter of the bounce period, $\omega \sim (2 \pi)/(0.5 \times 4) = \pi \; {\rm radian \; s^{-1}}$.
 Assuming that the magnetic field strength at the emission site is on the order of 100 Gauss (by Y2002), the velocity of non-thermal {\gyros}-emitting electrons is almost the speed of light \citep{1999spro.proc..211B}, $v \sim c = 3 \times 10^{5} \; {\rm km \; s^{-1}} $. Using these, we can evaluate the magnetic mirror ratio as well as the initial pitch-angle cosine from equations (\ref{eq:rr}) and (\ref{eq:omega}),
\begin{eqnarray}
\mu_{0} &=& \frac{\omega R}{c} \sim 0.47, \label{eq:mu0} \\
M &=& 1 + \frac{\omega r_{\rm f}/c}{1-\mu_{0}^{2}} \sim 1.3. \label{eq:mratio}  
\end{eqnarray}
Y2002 obtained a magnetic field strength at the loop top, $B_{0} \sim 200 \; {\rm Gauss}$ by potential field extrapolation. The photospheric magnetic field strength at the loop end is $\sim 400 \; {\rm Gauss}$.
We consider that the electron propagation end is above the photosphere where the field strength is probably lower than that at the photosphere, because numerous collisions with ambient plasma prevent them from reaching the photosphere.
 Therefore we consider $M \lsim 2$.
From this inequality and equation (\ref{eq:omega}), we obtain $\mu_{0} \lsim 0.88$. Consequently we evaluate $0.47 \lsim \mu_{0} \lsim 0.88$, thus $28 ^{\circ} \lsim \alpha_{0} \lsim 62^{\circ}$ which is smaller than that derived by Y2002. This is because we consider the field convergence while they did not. 

\section{Numerical Model}\label{sec:numerical-model}
 In the previous section we only treat the motion of a single electron. 
Our goal is to numerically model the {\gyros} intensity distribution along the loop for comparison with the observation. For this purpose, we first numerically model the electron motion along the loop by the {\FP} approach of the electron distribution in phase space.
 The gyro-averaged {\FP} equation is as follows \citep[][]{1988ApJ...327..405L,1990ApJ...354..726H,1990A&A...234..487M}:
\begin{eqnarray}
\pdif{N}{t} + \mu c \beta \pdif{N}{r} + \pdif{}{\mu}\left(\dot{\mu} N\right) + \pdif{}{E}\left(\dot{E} N\right) = \pdif{}{\mu}\left(D_{\mu \mu} \pdif{N}{\mu} \right). \label{eq:fp}
\end{eqnarray}
Here $N(r,\mu,E,t)$ is the electron distribution in phase space (number of electrons per unit length per unit pitch-angle cosine per unit energy), $\dot{E}$ and $D_{\mu \mu}$ are the Coulomb energy loss rate and pitch-angle diffusion coefficient, $\dot{\mu}$ is the magnetic mirroring force, $E = \Gamma - 1$ is the kinetic energy in units of the electron rest mass energy $m_{\rm e} {c}^2$, $\Gamma$ is the Lorentz factor, $\beta = \mart{1 - \Gamma^{-2}}$, $m_{\rm e}$ is the electron mass, respectively.

We adopt the Coulomb energy loss rate and pitch-angle diffusion coefficient for a fully ionized plasma given by \cite{1981ApJ...251..781L},
\begin{eqnarray}
&& \dot{E} = - K n / \beta \equiv - \nu_{\rm E} E, \;\;\; (K = 4 \pi c r_{0}^{2} \ln \Lambda), \label{eq:edot} \\
&& D_{\mu \mu} = \fr{K n}{\beta^{3} \Gamma^{2}} (1-\mu^{2}) = \frac{\nu_{\rm E}}{(E+2)}(1-\mu^{2}) ,\label{eq:dmumu}
\end{eqnarray}
where $r_{0} = 2.82 \times 10^{-13} \; {\rm cm}$ is the classical electron radius, $n$ is the ambient plasma number density, $\ln \Lambda$ is the Coulomb logarithm, and $\nu_{\rm E} \equiv K n/(\beta E)$ is the Coulomb collision frequency, respectively. The ambient plasma number density and the Coulomb logarithm are treated as constant in this paper, $n=10^{10} \; {\rm cm^{-3}}$ and $\ln \Lambda = 25$. Since we address shorter timescale (order of 1 s) phenomena than the Coulomb collision time (order of 100 s for 1 MeV electrons), the results and conclusion in this paper are almost independent of these variables.

\begin{figure}[htbp]
\centering
\plotone{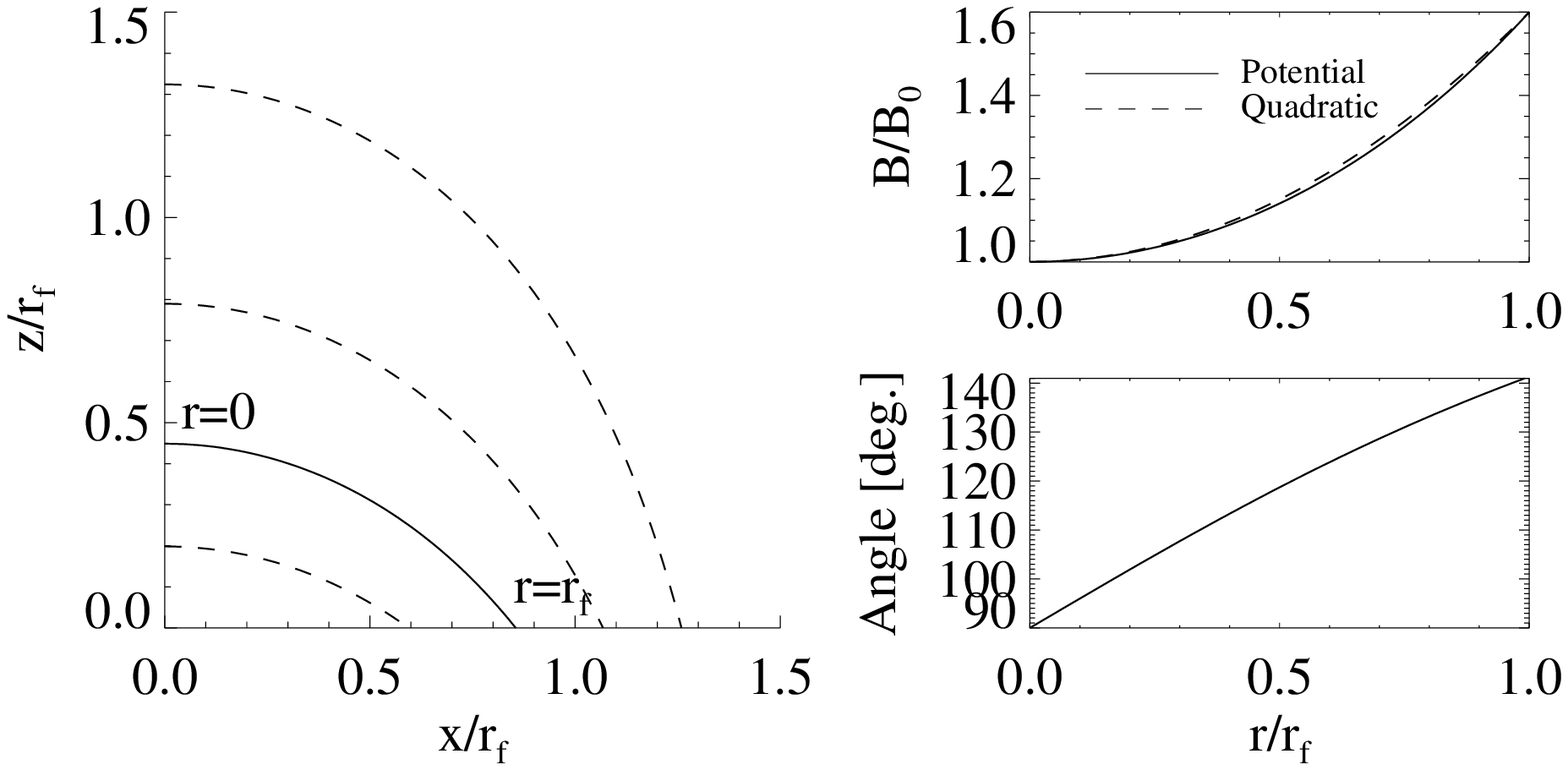}
\caption{{\it Left}: The potential magnetic field configuration described by equation (\ref{eq:pbx}). The $x$- and $z$-axes correspond to the tangential and normal directions relative to the solar surface, respectively. The $y$-axis is perpendicular to the plane. We solve the {\FP} equation (eq. (\ref{eq:fp})) along the thick line which has the half loop length $r_{\rm f} = 4.5 \times 10^{4} \; {\rm km}$ and the magnetic mirror ratio $M = 1.6$. {\it Top right}: The magnetic field strength measured along the field line (thick line in the left panel). The solid and dashed lines correspond to equations (\ref{eq:pbr}) and (\ref{eq:quadb}), respectively. {\it Bottom right}: A viewing angle $\theta$ with respect to the magnetic field line (thick line in the left panel) described by equation (\ref{eq:theta_disk}). The line of sight is assumed to be along the $z$-axis. When the line of sight is along the $y$-axis, $\theta = 90^{\circ}$ is independent of $r$.}
\label{fig:mag_strength}
\end{figure}

The magnetic mirroring force is given by
\begin{eqnarray}
\dot{\mu} = -\frac{1}{2} c \beta (1-\mu^{2}) \dif{\ln B(r)}{r}. \label{eq:mudot}
\end{eqnarray}
Knowledge of the field configuration is needed to give the strength $B(r)$ in this equation, and also to allow the simulation of the spatial distribution of the emission, because the {\gyros} radiation depends on a viewing angle with respect to the field line. 
We employ the two-dimensional, static, and symmetric potential magnetic field in Cartesian coordinates as follows \citep{1982QB539.M23P74...}:
\begin{eqnarray}
\left\{
\begin{array}{l}
B_{x} = B_{\rm f} \cos (kx) e^{-kz}, \label{eq:pbx}\\
B_{z} = - B_{\rm f} \sin (kx) e^{-kz},
\end{array}
\right.
\end{eqnarray} 
where $x$ and $z$ directions correspond to the tangential and normal directions relative to the solar surface, respectively. The parameter $k$ is determined from $r_{\rm f}$ and $M$ (see below). This field configuration is shown in the left panel of Figure \ref{fig:mag_strength}.
The strength of this field measured along a field line is written as
\begin{eqnarray}
B(r) = \frac{B_{0}}{\cos\left[\sin^{-1}\left\{\tanh(k r)\right\}\right]}, \label{eq:pbr}
\end{eqnarray}
where
\begin{eqnarray}
k = \frac{1}{2 r_{\rm f}} \ln \left[ \frac{1 + \sin\left\{\cos^{-1}\left(M^{-1}\right)\right\}}{1 - \sin\left\{\cos^{-1}\left(M^{-1}\right)\right\}} \right]. \label{eq:k_mratio}
\end{eqnarray}
The top right panel of Figure \ref{fig:mag_strength} shows the field strength measured along the field line. We use parameters $r_{\rm f} = 4.5 \times 10^{4} \; {\rm km}$ and $M = 1.6$ based on the estimation in {\S}~\ref{sec:analytic-treatment}. The form of the field strength (eq. (\ref{eq:pbr}), solid line in the top right panel of Fig. \ref{fig:mag_strength}) is very close to the quadratic function (eq. (\ref{eq:quadb}), dashed line), allowing us to use the analytic solution in {\S}~\ref{sec:analytic-treatment} as an approximation for the calculation result. 

We give the initial condition for $N$ as
\begin{eqnarray}
N(r,\mu,E,0) = \exp\left[-\left(\frac{r}{0.2r_{\rm f}}\right)^2\right] \phi(\mu) E^{-4}, \label{eq:ini_con}
\end{eqnarray}
where $\phi(\mu)$ is the initial pitch-angle distribution given later. By this equation, energetic electrons with a power-law distribution are initially injected at the loop top.
The initial condition is given to be symmetric in $(r, \mu)$ space, which yields the symmetric solution of equation (\ref{eq:fp}): $N(r,\mu,E,t) = N(-r,-\mu,E,t)$.

We numerically solve equation (\ref{eq:fp}) by using a finite difference method with operator splitting. The differential operators in equation (\ref{eq:fp}) are split into four terms: the advection terms in $r$, $\mu$, and $E$ space, and the diffusion term. We solve the diffusion term by using the Crank-Nicholson method with central difference. We employ the symmetric boundary condition in $\mu$ space to ensure total number conservation. For the advection term in $E$ space we use an analytic solution derived by the method of characteristics \citep{1985Ap&SS.116..377C,1986A&A...163..239M}, and adopt a single power-law function for the necessary interpolation at the intermediate location between the grid points (if either grid point has a non-positive value, we instead apply a linear interpolation). For the advection terms in $r$ and $\mu$ space, we adopt the CIP-CSL2 scheme developed by \cite{2001mwr...129..332Y}. This scheme simultaneously solves the integrated value of $N$ as well as $N$ itself to keep ``subgrid'' information and to satisfy the mass conservation. A boundary condition at $r=0$ is $\partial\left\{N(r,\mu,E,t) - N(r,-\mu,E,t)\right\}/\partial r = 0$,
for the symmetry. Another boundary condition at $r=r_{\rm f}$ is as follows: (1) electrons with positive pitch angle, $N(r_{\rm f},\mu \geq 0, E,t)$, are lost from the calculation box; and (2) there is no flow from outside the box, $N(r_{\rm f}, \mu < 0, E,t) = 0$. Physically, the former corresponds to the electron precipitation into the footpoint, and the latter means that the possibility of electrons originating below the footpoint (chromosphere) is excluded.
 We set $64 \times 128$ cells in ($r,\mu$) space, and 128 logarithmically-spaced grids from 50 to 5000 keV in $E$ space.

Using the electron distribution $N(r,\mu,E,t)$, we further numerically calculate the distribution of the {\gyros} radiation for comparison with the observation of Y2002. A general calculation of the {\gyros} radiation in a magnetized plasma includes effects such as self-absorption, absorption by ambient plasma, and the Razin suppression \citep{1969ApJ...158..753R}. These effects contribute at low harmonics ($\nu / \nu_{\rm B} \lsim 10$, where $\nu_{\rm B}$ is the electron gyrofrequency) of the radiation in the optically-thick regime. However, Y2002 showed that the 17 GHz emission observed with the NoRH in the 1999 August 28 flare is the optically-thin, non-thermal {\gyros} radiation. Therefore, it is sufficient for our purpose in this paper to consider only the optically-thin radiation at high harmonics ($10 \lsim \nu / \nu_{\rm B} \lsim 100$) from mildly relativistic electrons ($\Gamma \lsim 10$). Under such limited condition,  \cite{1981ApJ...251..727P} gave an useful approximation to calculate the {\gyros} emissivity from mildly relativistic electrons with arbitrary energy and pitch-angle distributions (eq. (5) in his paper). We adopt his formula to calculate the radiation from 17 to 34 GHz.

The {\gyros} emissivity $j_{\nu}(r,t)$ depends on a magnetic field strength and a viewing angle $\theta$ with respect to the field line at the emission site as well as the parent electron distribution. We set a field strength at the loop top of $B_{0} = 200$ Gauss. 
A viewing angle, depending on the loop location, orientation, and tilt with respect to the Sun, is calculated from the direction of the field line and the line of sight. We consider two ideal cases of the viewing angle. When the line of sight is along the $z$-axis (see the left panel of Fig. \ref{fig:mag_strength}), $\theta$ is given as a function of $r$,
\begin{eqnarray}
\theta(r) = \cos^{-1}\left[ -\tanh(k r)\right], \label{eq:theta_disk} 
\end{eqnarray}
that is shown in the bottom right panel of Figure \ref{fig:mag_strength}. We call this as the top-view case. Such situation is realized when the loop is at the disk center and is not tilted with respect to the solar surface.
When the line of sight is along the $y$-axis, $\theta = 90^{\circ}$ is independent of $r$. We call this as the side-view case. Such situation is realized when the loop is on the limb with north-south orientation.


We calculate the {\gyros} intensity $I_{\nu}(r,t)$, which is the observed variable, from the emissivity for comparison with the observation. This can be obtained by considering the projection effect of the loop by an observer,
\begin{eqnarray}
I_{\nu}(r,t) = j_{\nu}(r,t) \frac{\mart{A(r)}}{\sin \theta}, \label{eq:j2i}
\end{eqnarray}
where $A(r)$ is the cross-sectional area of the loop determined from the conservation of the magnetic flux, $B(r)A(r) = {\rm constant}$.

\section{Calculation Results and Interpretation}\label{sec:result-discussion}
We present our model calculation results and address whether the apparent propagating motion of the microwave source reported by Y2002 actually corresponds to the motion of electrons injected with a specific initial pitch angle. To discuss this, we perform calculations for two cases of the pitch-angle distribution of the initial condition ($\phi(\mu)$ in eq. (\ref{eq:ini_con})): narrow-angle injection ({\S}~\ref{sec:narrow-band-case}) and broad-angle injection ({\S}~\ref{sec:broad-band-case}) cases.

\subsection{Narrow-Angle Injection}\label{sec:narrow-band-case}
We first consider that initially injected electrons have an almost unique pitch angle, following the interpretation made by Y2002. 
To simulate this case, we give an initial pitch-angle distribution peaking at $\mu = \pm \mu_{\rm p}$:
\begin{eqnarray}
\phi(\mu) = \exp \left[ -\left(\frac{|\mu|-\mu_{\rm p}}{0.1}\right)^{2}\right]. \label{eq:ini_mu_narrow}
\end{eqnarray}

\begin{figure}[htbp]
\centering
\plotone{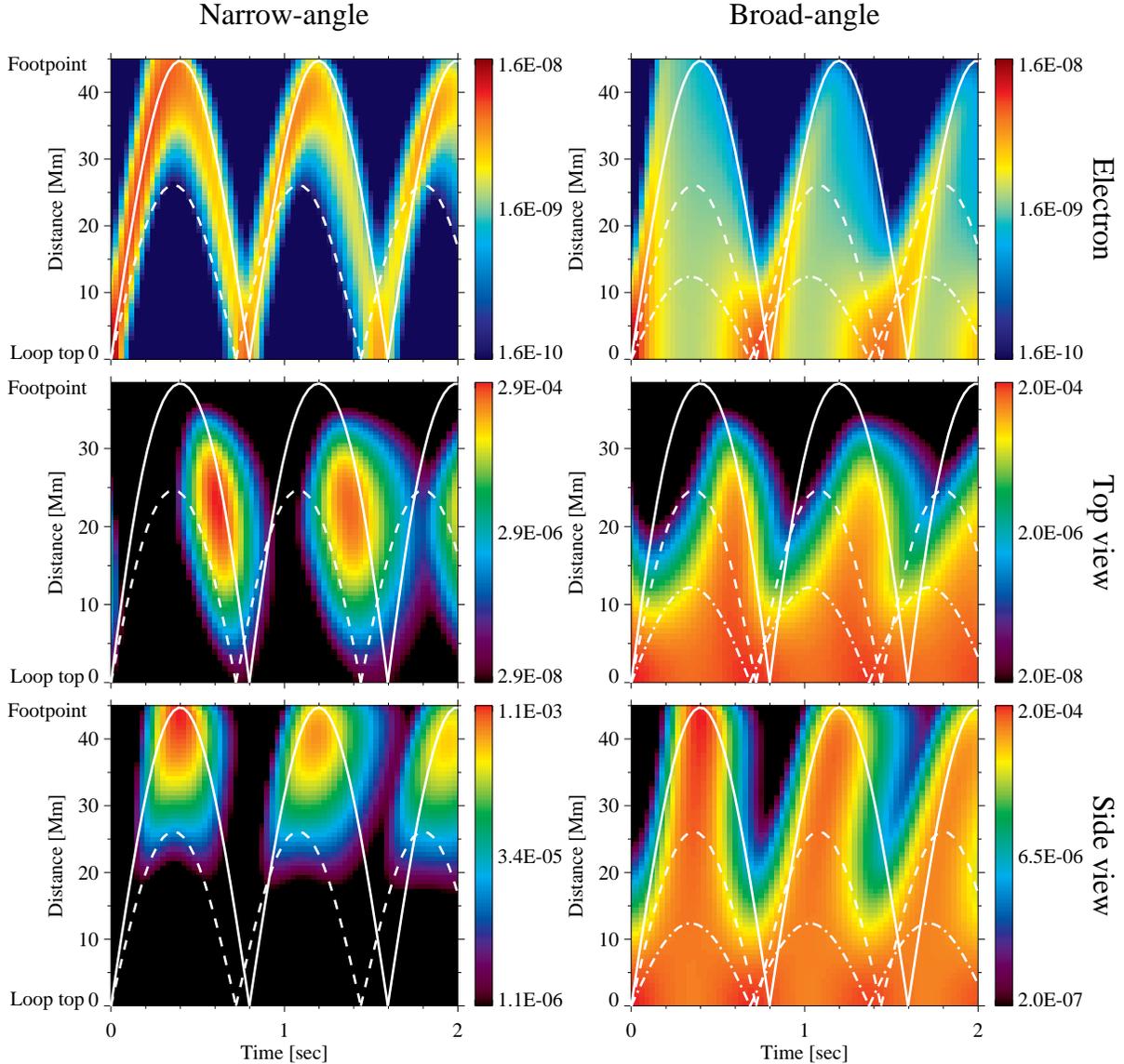}
\caption{Time variations of the electron number ({\it top}) and the {\gyros} intensity ({\it middle} and {\it bottom}) distributions along the loop. The vertical axis corresponds to the spatial coordinates from the loop top to the footpoint. The {\it left} and {\it right} panels correspond to the results calculated for the narrow- and broad-angle injection cases, respectively. {\it Top}: The 1 MeV electron number distribution along the loop. {Relativistic electrons show almost the same distribution as that of 1 MeV electrons.} {\it Middle}: The 17 GHz intensity distribution along the loop in the top view. Note that the vertical scale of these panels is slightly different from other panels because of the projection effect of the loop. {\it Bottom}: The 17 GHz intensity distribution along the loop in the side view. The solid, dashed, and dash-dotted lines show the trajectories of electrons with initial pitch-angle cosine of $\pm 0.61$, $\pm 0.4$, and $\pm 0.2$, respectively.}
\label{fig:result}
\end{figure}

The top left panel of Figure \ref{fig:result} shows the time variation of the electron number distribution along the loop.
We show the calculation result with $\mu_{\rm p} = \mart{1-M^{-1}} = 0.612$, which is equal to the loss cone angle cosine $\mu_{\rm c}$.
The vertical axis corresponds to the spatial coordinates from the loop top to the footpoint. 
We use the 1 MeV electron distribution integrated over pitch angle for this illustration.
The solid and dashed lines show the trajectories of electrons with initial pitch-angle cosine of respective $\pm 0.61$ and $\pm 0.4$, obtained by solving equation (\ref{eq:drdt}) with equations (\ref{eq:mag_mom}) and (\ref{eq:pbr}) (not eq. (\ref{eq:quadb})) by using the 4th order Runge-Kutta method. 

Initially injected electrons move toward the footpoint and about a half of them reach there after 0.2 - 0.4 s. They are lost from the calculation box because they are initially in the loss cone. Remaining electrons bounce back by the magnetic mirror and move toward the opposite direction. 
Subsequent diffusion by Coulomb collisions is ignorable because the collision time  ($\sim 200$ s) is much longer than the loop transit time ($< 1$ s).

{
We calculate the electron distribution not only in space but also in energy (see eqs. (\ref{eq:fp}) and (\ref{eq:ini_con})). Since the electron propagation is velocity-dispersive, the time variation of the electron number distribution differs with energy. 
However, relativistic electrons show almost the same profile as the top left panel of Figure \ref{fig:result} because their velocity is nearly $c$.
}

The middle and bottom left panels of Figure \ref{fig:result} show the time variation of the 17 GHz {\gyros} intensity distribution along the loop in the top- and side-view cases, respectively. Note that the vertical scale of the middle left panel is slightly different from other panels because of the projection effect of the loop.


The strong radiation source is localized in space along the loop. In the side view (bottom left), radiation primarily comes from the footpoint. In the top view (middle left), strong radiation comes from the intermediate position between the loop top and footpoint. These results are attributed to the dependence of the {\gyros} emissivity on a viewing angle as well as the parent electron distribution along the loop. 
We interpret these results by using the analytic solution in {\S} \ref{sec:analytic-treatment}.

The {\gyros} radiation is primarily produced by mildly relativistic ($\sim 1$ MeV) electrons with the pitch-angle cosine of $\mu \sim \beta \cos \theta$ \citep{1981ApJ...251..727P,1990ApJ...354..735L}.
Given all electrons with the initial pitch-angle cosine $\mu_{0}$ at the loop top, their pitch-angle cosine is given by equation (\ref{eq:mag_mom}) with (\ref{eq:quadb}) as a function of $r$.
The viewing angle $\theta$ is also given as a function of $r$, which is different between the top- (eq.~(\ref{eq:theta_disk})) and side-view ($\theta = 90^{\circ}$) cases. By solving $\mu = \beta \cos \theta$ with respect to $r$, we obtain the position $r_{\rm g}$ at which strong radiation are produced.
In the top view, the equation is written as
\begin{eqnarray}
\mart{\mu_{0}^{2} - (1-\mu_{0}^2)(M-1)\left(\frac{r_{\rm g}}{r_{\rm f}}\right)^{2}} = \beta \tanh (k r_{\rm g}). \label{eq:rg_disk}
\end{eqnarray}
The solution $r_{\rm g}$ of equation (\ref{eq:rg_disk}) is smaller than $r_{\rm f}$, as long as $\mu_0$ is smaller than the critical value (obtained by taking $r_{\rm g} = r_{\rm f}$ in this equation). 
In the narrow-angle injection case, we assume that injected electrons have an almost unique initial pitch angle, $\mu_{0} \sim \mu_{\rm p}$.
When $\mu_{\rm p} = \mart{1-M^{-1}}$, equation (\ref{eq:rg_disk}) is 
\begin{eqnarray}
\beta \tanh (k r_{\rm g}) \sim \mart{\left(1-M^{-1}\right)\left\{1-\left(\frac{r_{\rm g}}{r_{\rm f}}\right)^{2}\right\}},\label{eq:rg_disk_rewrite}
\end{eqnarray}
which gives $r_{\rm g}$ of an intermediate value between 0 and $r_{\rm f}$. Therefore the strong radiation comes from the intermediate position between the loop top and footpoint. This explains the intensity distribution in the middle left panel of Figure \ref{fig:result}.

In the side view, $\theta = 90^{\circ}$ is independent of $r$. Then the equation $\mu = \beta \cos \theta = 0$ results in
\begin{eqnarray}
r_{\rm g} = \frac{\mu_{0}}{\mart{(1-\mu_{0}^{2})(M-1)}} r_{\rm f},\label{eq:rg_limb}
\end{eqnarray}
which again gives $r_{\rm g}$ of an intermediate value between 0 and $r_{\rm f}$ as long as $\mu_{0} < \mu_{\rm c} = \mart{1-M^{-1}}$. 
When $\mu_{0} \sim \mu_{\rm p} = \mart{1-M^{-1}}$, this results in $r_{\rm g} \sim r_{\rm f}$.
Therefore the strong radiation comes from the footpoint. This explains the intensity distribution in the bottom left panel of Figure \ref{fig:result}.

Based on these discussions, we conclude that electrons injected into the loop with an almost unique pitch angle do not yield the propagating feature of the radiation source along the loop.

\subsection{Broad-Angle Injection}\label{sec:broad-band-case} 
Next, we consider that initially injected electrons have an isotropic pitch-angle distribution: $\phi(\mu) = {\rm constant}$. Electrons with a small initial pitch angle can reach the footpoint while those with a large initial pitch angle are confined to a narrow region around the loop top.

The top right panel of Figure \ref{fig:result} shows the time variation of the electron number distribution along the loop. The solid, dashed, and dash-dotted lines show the trajectories of electrons with initial pitch-angle cosine of respective $\pm 0.61$, $\pm 0.4$, and $\pm 0.2$. As expected, electrons are broadly distributed in the loop compared with the narrow-angle injection case (top left panel).

The middle and bottom right panels of Figure \ref{fig:result} show the time variation of the 17 GHz {\gyros} intensity distribution along the loop in the top- and side-view cases, respectively. The intensity distribution is quite different from that in the narrow-angle injection case (middle and bottom left panels). The strong radiation source is broadly distributed along the loop. We can see the propagating feature of the strong radiation source from the loop top to the footpoint in both the middle and bottom right panels, similar to the observation of Y2002. For example, rapidly propagating features are found during 0.3 - 0.5 s and 0.9 - 1.2 s in the bottom right panel. These features do not result from the motion of electrons with a specific initial pitch angle but from the motion of an ensemble of electrons with different initial pitch angles.

\begin{figure}[htbp]
\centering
\epsscale{0.8}
\plotone{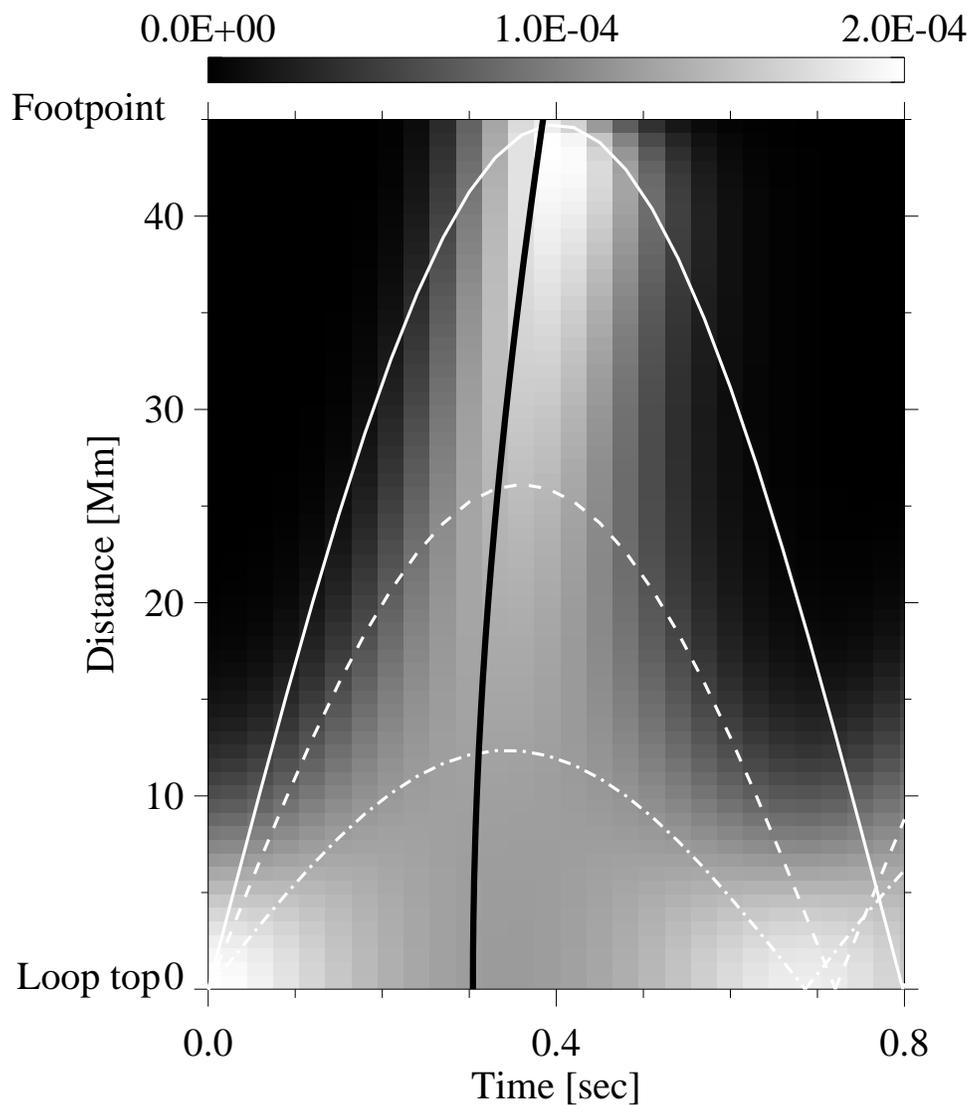}
\caption{Time variation of the 17 GHz {\gyros} intensity distribution along the loop in the side view during 0-0.8 s, calculated for the broad-angle injection case. The solid, dashed, and dash-dotted lines show the trajectories of electrons with initial pitch-angle cosine of $\pm 0.61$, $\pm 0.4$, and $\pm 0.2$, respectively. The thick black line shows the analytic solution (eq. (\ref{eq:ana_solution})).}
\label{fig:result_zoom}
\end{figure}

We discuss the apparent motion of the radiation source by using the analytic solution in {\S} \ref{sec:analytic-treatment}. For mathematical simplicity, we consider the side-view case. As mentioned in {\S}~\ref{sec:narrow-band-case}, the {\gyros} radiation is primarily produced by the electrons with $\mu \sim 0$ in the side view.
It is found from equations (\ref{eq:sol_mu}) and (\ref{eq:omega}) that the pitch-angle cosine of an electron with the initial pitch-angle cosine $\mu_{0}$ becomes zero at
\begin{eqnarray}
t_{\rm g} = \frac{(m/2) \pi}{\omega} = \frac{(m/2)\pi}{\mart{(1-\mu_{0}^{2})(M-1)}} \frac{r_{\rm f}}{v}, \label{eq:t_g}
\end{eqnarray}
where $m$ is an odd integer. This equation shows that when $\mu_{0}$ is smaller (i.e., an initial pitch angle is larger) $t_{\rm g}$ is smaller. At $t = t_{\rm g}$, the electron position is given by equation (\ref{eq:sol_r}) with $\sin(\omega t_{\rm g}) = 1$. 
Using this, $\mu_{0}$ in equation (\ref{eq:t_g}) can be removed and it is rewritten as
\begin{eqnarray}
t_{\rm g} = \frac{m \pi}{2} \mart{1 + \frac{1}{M-1}\left(\frac{r_{\rm f}}{r_{\rm g}}\right)^2} \frac{r_{\rm g}}{v}. \label{eq:ana_solution}
\end{eqnarray}
This equation shows that when $r_{\rm g}$ is smaller $t_{\rm g}$ is smaller. Therefore, $r_{\rm g}$ is smaller when $\mu_{0}$ is smaller.
Electrons produce strong radiation at different time and position, depending on their initial pitch angle.
By this equation, one can trace the position of the strong {\gyros}-emitting electron in $(r,t)$ space.
 Figure \ref{fig:result_zoom} is the zoomed image of the bottom right panel of Figure \ref{fig:result}, to compare the analytic solution (eq. (\ref{eq:ana_solution})) with the calculation result. The thick black line shows the analytic solution with $m=1$ and $v=c$. As can be seen, the analytic solution well explains the propagating feature of the radiation source. 

Based on these discussions, we interpret the microwave propagating feature as follows. Electrons with a larger (smaller) initial pitch angle emit microwaves earlier (later) at the position closer to (farther from) the loop top, as seen in the peak of the dash-dotted (solid) line in Figure \ref{fig:result_zoom}. This difference, due to the difference of the initial pitch angle of the parent electrons, appears as the propagating motion of the strong microwave source along the loop from the loop top to the footpoint.

 The propagation of the radiation source seen in Figure \ref{fig:result} is periodic. 
This is because the initially injected electrons remain to oscillate in the loop $(m=1,3,5,\dots)$.
Actual flares, on the other hand, show the light curve with a duration (tens of second) longer than the electron loop transit time scale ($\lsim 1$ s), indicating that electrons are continuously injected into the loop \cite[e.g.,][]{2008ApJ...673..598M}.
The periodic propagation seen in our calculation is probably occulted by the emissions from subsequently injected electrons, except the first one. 
The observable propagating motion is the first line $(m=1)$.
The time of electrons reaching the footpoint and the propagation speed of the radiation source are calculated from equation (\ref{eq:ana_solution}) with $r_{\rm g} = r_{\rm f}$,
\begin{eqnarray}
t_{\rm end} = \frac{\pi}{2}\mart{\frac{M}{M-1}} \frac{r_{\rm f}}{v},\;
v_{\rm prop} \simeq \frac{r_{\rm f}}{t_{\rm end}} = \frac{2}{\pi}\mart{\frac{M-1}{M}}v. \label{eq:vprop}
\end{eqnarray}
{
The propagation speed of the radiation source is determined from the electron velocity and the magnetic mirror ratio. When the magnetic field strength at the emission site is on the order of 100 Gauss, effective energy of electrons producing $\gsim 10 \;{\rm GHz}$ {\gyros} radiation is $\gsim 1 $MeV \citep{1999spro.proc..211B}, hence their velocity is nearly $c$. 
The propagation speed, $v_{\rm prop} \simeq 0.4c$ with $M=1.6$, is almost independent of the observation frequency, as long as parent electrons are relativistic. This frequency-independent microwave propagating feature is found not only at 17 GHz but also at higher frequencies up to 34 GHz in our calculation.
}


We performed the calculation for two ideal cases of the view (top and side) because the {\gyros} radiation depends on the viewing angle. The apparent propagating motion seen in the top-view case (middle right panel of Fig. \ref{fig:result}) looks different from that in the side-view case (bottom right panel), but can qualitatively be understood in the same manner as the side view. Difference of the view, which depends on the magnetic field configuration and location of the loop, does not influence our interpretation on the microwave propagating feature.


It is supposed that a specific electron accounts for the propagating feature of the microwave source. 
To produce the strong radiation, the pitch angle $\alpha$ of the electron should be close to $\theta$ at any time and position throughout its propagation along the loop.
This is satisfied when (1) both $\alpha$ and $\theta$ do not change, or (2) $\theta$ changes in the same manner as $\alpha$. The former case is realized if the field strength and the viewing angle do not change along the loop. The latter case may be realized if the loop is on the limb with east-west orientation, because $\theta$ as well as $\alpha$ increase toward the footpoint in such configuration.
These cases are, however, limited ones. Our interpretation holds in any probable field configuration and loop location.


\section{Summary and Discussion}\label{sec:summary}
We presented analytic and numerical treatments of the electron motion along a magnetic loop, and consider the rapidly propagating feature of the non-thermal microwave source reported by Y2002.
 We studied this issue by describing the electron distribution with the {\FP} equation and by calculating the {\gyros} radiation.

We first assumed that electrons injected into the loop have an almost unique initial pitch angle, following the interpretation made by Y2002. These electrons do not yield the propagating feature of the radiation source along the loop. This does not support the interpretation of Y2002.

We next assumed that electrons injected into the loop have an isotropic pitch-angle distribution. In this case, the intensity distribution shows the apparent motion of the strong radiation source from the loop top to the footpoint, similar to the observation. This feature is interpreted as the motion of an ensemble of electrons (not a specific electron), which have different time and position to produce strong radiation due to the difference of their initial pitch angle. To show the microwave propagating feature, the injected electrons should be broadly distributed in pitch-angle space.

We discuss a probable physical process of electrons in the 1999 August 28 flare. It is thought that this flare was triggered by the interaction between the microwave-propagating loop and a compact loop. Such configuration is suggested by \cite{1999PASJ...51..483H}, called ``double-loop flare''. \cite{1999PASJ...51..483H} concluded that in the double-loop flare electrons are accelerated at the region where two loops interact. Based on this model, Y2002 interpreted that the acceleration site in the 1999 August 28 flare is where the propagation of the non-thermal microwave source starts, that is, the injection region. We conclude that the electrons in the 1999 August 28 flare started to propagate along the loop just after being isotropically accelerated at the site.

In {\S} \ref{sec:narrow-band-case} we discussed the calculation for the narrow-angle injection case with $\mu_{\rm p} = 0.612$ only. The discussion can be applied to electrons with different pitch-angle distributions. When the electron pitch-angle distribution is concentrated perpendicular to the magnetic field, $\mu_{\rm p} = 0$. In this case, it is found from equations (\ref{eq:rg_disk}) or (\ref{eq:rg_limb}) with $\mu_0 \sim 0$ that the strong radiation source is confined to the loop top. When the electron pitch-angle distribution is concentrated parallel to the field, on the other hand, $\mu_{\rm p} = 1$. In this case the solution of equations (\ref{eq:rg_disk}) or (\ref{eq:rg_limb}) with $\mu_0 \sim 1$ is $r_{\rm g} = \infty$, meaning that there is no position within the loop to satisfy the condition for strong radiation. Neither of these cases results in the microwave propagation.

{We calculated the {\gyros} intensity from the approximation of \cite{1981ApJ...251..727P}, which is valid when ambient plasma density at the emission site is so low that the absorptions and the Razin suppression are not important. We have to use a more exact formula of the {\gyros} intensity \citep[e.g.,][]{1969ApJ...158..753R,2003ApJ...587..823F} in case that microwaves are produced at the high-density site in which these effects are important.}

{As well as the intensity, one can utilize the degree of circular polarization of the {\gyros} radiation for diagnostics of the pitch-angle distribution of parent electrons \citep[e.g.,][]{2008ApJ...677.1367A}. \cite{2003ApJ...587..823F} found from their numerical calculations that the degree of polarization increases when the electron pitch-angle distribution is more anisotropic and/or the line of sight is more parallel to rather than perpendicular to the magnetic field line. Propagating electrons in a converging magnetic loop have a more anisotropic distribution at the footpoint than the loop top. Furthermore, it is expected for the disk flares (top view) that the magnetic field line at the footpoint is close to parallel to the line of sight.
 Therefore, {\gyros} emission is more likely to be polarized at the footpoint than the top of the electron-propagating loop. 
The 1999 August 28 flare certainly showed that the degree of polarization increases toward the footpoints from the loop top (see Fig. 2 in Y2002), agrees with the above statement.}

The propagating feature of the microwave source gives us great opportunities to study the {\gyros} radiation mechanism and the electron transport, and to constrain the pitch-angle distribution of the injected electrons which is crucially important to understand the electron acceleration mechanism. As is evident from many hard X-ray observations showing footpoint sources \citep[e.g.,][]{1994PhDT.......335S,1999spro.proc..321S}, propagation of electrons and the microwave source along the loop commonly occurs in the flare. However, detection of such phenomena is quite difficult. 
The 1999 August 28 flare is the unique event in that NoRH detected the microwave propagating feature during its observational period until end of 2004 since its operation start in 1992 June \citep{Shimojo_priv}. This event was an extremely well resolved one in both space and time by NoRH.

 For further study, improvement of radio observatories is important. Observations with high temporal ($\lsim 0.1 \; {\rm s}$) and spatial ($\sim 1''$) resolutions more clearly resolve the microwave source because it propagates with a speed close to the speed of light along a loop with typical length $\lsim 100'' \sim 7 \times 10^{4} \; {\rm km}$. Such observations should be implemented at frequencies greater than $\sim 10 \; {\rm GHz}$ which correspond to the optically-thin regime of the microwave emission in typical solar flares, to obtain the spectral property of non-thermal electrons.
Since the {\gyros} radiation depends on the viewing angle with respect to the loop, the center-to-limb variation of the microwave distribution is studied to address the electron distribution in the loop. Such study has been carried out statistically by e.g., \cite{1985PASJ...37..575K} and \cite{2002SoPh..206..177S} with spatially-unresolved data. Statistical study on the center-to-limb variation of the spatial distribution of the microwave emission further gives constraints on the pitch-angle distribution of the injection electrons. Our numerical study will be of help to the future observational study for understanding the electron dynamics in solar flares.

\begin{acknowledgements}
We would like to thank M. Hoshino, K. Makishima, M. Fujimoto, S. Masuda, and S. Tsuneta for fruitful discussions and variable comments. We also thank K. Shibasaki, M. Shimojo, A. Asai, and H. Nakajima for discussions and advises on the NoRH data. We thank the anonymous referee's helpful comments and corrections to our manuscript.
One of the authors (T. M.) was supported by COE program of the University of Tokyo, ``Predictability of the Evolution and Variation of the Multi-scale Earth System: An integrated COE for Observational and Computational Earth Science''.
\end{acknowledgements}



\end{document}